\documentclass[twocolumn,superscriptaddress,PRB]{revtex4}

\usepackage{url}

\usepackage{amsmath}
\usepackage{amssymb}
\usepackage{graphicx}

\begin{document}

\title{Imaging two-component nature of Dirac-Landau levels in the topological surface state of Bi$_2$Se$_3$}

\author{Ying-Shuang Fu}
\email{yfu@riken.jp}
\affiliation{RIKEN Center for Emergent Matter Science, Wako, Saitama 351-0198, Japan}

\author{M. Kawamura}
\email{minoru@riken.jp}
\affiliation{RIKEN Center for Emergent Matter Science, Wako, Saitama 351-0198, Japan}

\author{K. Igarashi}
\affiliation{Materials and Structures Laboratory, Tokyo Institute of Technology, Yokohama, Kanagawa 226-8503, Japan}

\author{H. Takagi}
\affiliation{Magnetic Materials Laboratory, RIKEN, Wako, Saitama 351-0198, Japan}
\affiliation{Department of Physics, University of Tokyo, Hongo, Bunkyo-ku, Tokyo 113-0033, Japan}
\affiliation{Max-Planck-Institut f\"ur Festk\"orperforschung, Heisenbergstra{\ss}e 1, 70569 Stuttgart, Germany}

\author{T. Hanaguri}
\email{hanaguri@riken.jp}
\affiliation{RIKEN Center for Emergent Matter Science, Wako, Saitama 351-0198, Japan}

\author{T. Sasagawa}
\affiliation{Materials and Structures Laboratory, Tokyo Institute of Technology, Yokohama, Kanagawa 226-8503, Japan}

\maketitle

\textbf{
Massless Dirac electrons in condensed matter have attracted considerable attention~\cite{Geim2007NatMat, CastroNeto2009RMP, Tajima2007EPL, Ran2009PRB, Hasan2010RMP, Qi2011RMP}.
Unlike conventional electrons, Dirac electrons are described in the form of two-component wave functions.
In the surface state of topological insulators, these two components are associated with the spin degrees of freedom~\cite{Hasan2010RMP, Qi2011RMP}, hence governing the magnetic properties.
Therefore, the observation of the two-component wave function provides a useful clue for exploring the novel spin phenomena.
Here we show that the two-component nature is manifested in the Landau levels (LLs) whose degeneracy is lifted by a Coulomb potential.
Using spectroscopic-imaging scanning tunneling microscopy, we visualize energy and spatial structures of LLs in a topological insulator Bi$_2$Se$_3$.
The observed potential-induced LL splitting and internal structures of Landau orbits are distinct from those in a conventional electron system~\cite{Hashimoto2012PRL} and are well reproduced by a two-component model Dirac Hamiltonian.
Our model further predicts non-trivial energy-dependent spin-magnetization textures in a potential variation.
This provides a way to manipulate spins in the topological surface state.
}

Landau quantization associated with the quasi-classical cyclotron motion of electrons is a fundamental phenomenon in a magnetic field $B$ and highlights the difference between conventional and Dirac electrons.
In conventional systems, the energy of $n$-th LL (LL$_n$), $E_n$, is proportional to $(n+\gamma)B$, where $\gamma=1/2$.
Distinct from this, $E_n$ in two-dimensional massless Dirac systems behave as $\propto \sqrt{|n|B}$~\cite{McClure1956PR,Goerbig2011RMP}.
Importantly, the Berry-phase effect in Dirac systems eliminates $\gamma$ and ensures the $B$-independence of $E_0$ which is equal to the Dirac-point energy~\cite{Novoselov2005Nature,Zhang2005Nature}.
Such an unusual LL sequence has been observed by scanning tunneling microscopy and spectroscopy (STM/STS) in graphene~\cite{Li2009PRL, Miller2009Science} and in the topological surface state~\cite{Hanaguri2010PRB,Cheng2010PRL}.

In addition to the unique energy spectrum, the wave functions of Dirac LLs are remarkably different from their conventional counterparts because of the two-component nature~\cite{Goerbig2011RMP}.
To study the details of wave functions, spectroscopic-imaging STM (SI-STM) is a powerful technique because tunneling-conductance maps, which include the information of the internal structures of wave functions through local-density-of-states (LDOS) variations, can be obtained with high energy and spatial resolutions.
If the system is uniform, the spatial degeneracy of Landau orbits results in a homogeneous LDOS.
The introduction of a potential variation lifts the spatial degeneracy, making it possible to access the localized Landau orbit~\cite{Morgenstern2003PRL,Hashimoto2008PRL,Miller2010NatPhys,Niimi2006PRL}.
The Landau orbit drifts along the equipotential lines and the LDOS variation across the orbit contains information of the internal structure of the wave function.
Indeed, a recent SI-STM study on a conventional two-dimensional electron system revealed the $n$-dependent nodal structure in the wave function~\cite{Hashimoto2012PRL}.

The wave-function imaging could be even more interesting in Dirac systems, because a potential variation not only lifts the spatial degeneracy of Landau orbits but also may affect the interplay between the two components in the wave function.
This is particularly important for the topological surface state where the interplay determines the magnetic properties.
Thus, exploring the two-component nature by the wave-function imaging will give us a clue for developing a novel spin-manipulation protocol.
For this, we study LL wave functions of a prototypical topological insulator Bi$_2$Se$_3$ using SI-STM.

\begin{figure*}
\includegraphics[width=120mm]{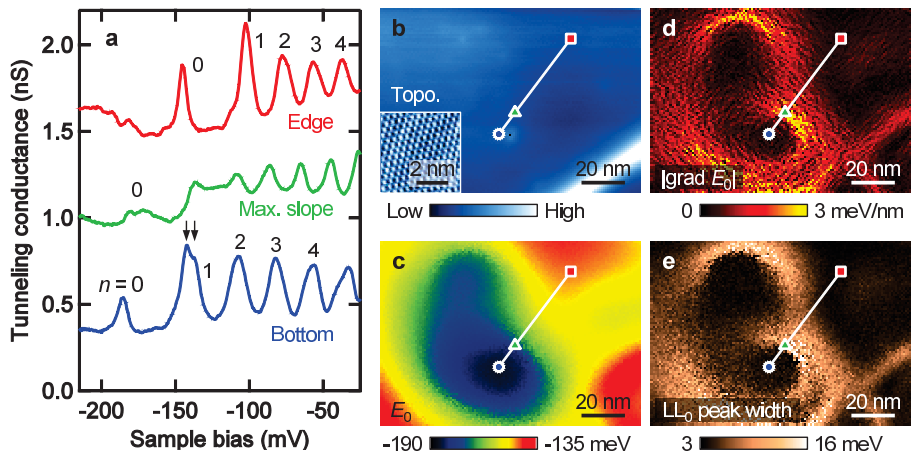}
\caption{
\textbf{Landau-level spectra and potential landscape in the topological surface state of Bi$_\textbf{2}$Se$_\textbf{3}$.}
\textbf{a},
Tunneling spectra taken at representative points along the line indicated in \textbf{b-e}.
Blue, green and red curves (from bottom to top) denote the data taken at the potential minimum (blue filled circle), at the potential-gradient maximum (green filled triangle) and near the edge of the potential dip (red filled square), respectively.
Each curve is offset vertically for clarity.
Data were taken at 1.5~K with conditions of sample-bias voltage $V_s=+50$~mV, tunneling current $I_t=50$~pA and bias modulation amplitude $V_{\rm mod}=2.1$~mV$_{\rm rms}$.
Note that the apparent LL$_0$ peak consists of a few peaks at the potential-gradient maximum and the LL$_1$ peak at the potential minimum splits into two peaks (black arrows).
\textbf{b},
Constant-current STM topograph of the cleaved surface.
Inset shows the atomic resolution image obtained by scanning the small area.
\textbf{c},
Potential landscape of the same field of view obtained by mapping $E_0$.
\textbf{d},
Potential-gradient map calculated from \textbf{c}.
\textbf{e},
Map of the apparent width (half width at half maximum) of LL$_0$ peak.
These four images (\textbf{b-e}) were taken simultaneously with $V_s=+50$~mV, $I_t=50$~pA and $V_{\rm mod}=2.8$~mV$_{\rm rms}$.
(For inset of \textbf{b}: $V_s=-100$~mV, $I_t=50$~pA)
The LL$_0$ peak in the individual spectrum was fitted with a single Lorentzian function to obtain $E_0$ and the apparent width of the peak.
}
\end{figure*}

Figure~1a represents LL spectra at $B=11$~T taken at marked points in the topographic image shown in Fig.~1b.
We confirm that $E_n$ exhibits $B$ and $n$ dependence typical for Dirac electrons~\cite{Hanaguri2010PRB} (see Supplementary Information).
The potential landscape can be visualized by mapping the spatial variation of $E_0$.
Since LL$_0$ is $B$ independent and is located at the Dirac-point energy, the $E_0$ map faithfully represents the potential landscape, albeit it is smeared over the size of the LL$_0$ wave function given by the magnetic length $l_B=\sqrt{\hbar/(|e|B)}$.
Here, $\hbar$ is the Planck constant divided by 2$\pi$ and $e$ is the elementary charge.
At 11~T, $l_B$ is $\sim7.7$~nm.
As shown in Fig.~1c, there is a well-defined potential minimum in the field of view.
This potential minimum may be generated by sub-surface charged defects (such as Se vacancies) because the topographic image (Fig.~1b) does not exhibit obvious defects at the surface.
Potential variations with similar length scale were also observed in graphene~\cite{Zhang2009NatPhys} and doped topological insulators~\cite{Beidenkopf2011NatPhys}.

We find that the potential-gradient map (Fig.~1d) exhibits strong correlation with the map of the apparent width of the LL$_0$ peak (Fig.~1e), implying that the spatial variation of potential lifts the degeneracy of LLs.
This is clearly manifested in the individual tunneling-conductance spectra shown in Fig.~1a.
The LL$_0$ peaks are sharp and single peaks at the potential minimum (blue) and at the 
edge of the potential dip where the potential becomes almost flat (red).
At the potential-gradient maximum (green), the LL$_0$ peak is not simply broadened but splits into multiple peaks which correspond to different quantum states as described later.
Recently, similar splitting has also been observed in graphene~\cite{Luican-Mayer2014PRL}.
Interestingly, the LL$_1$ peak splits into two peaks even at the potential minimum.
We will show below that this splitting of the LL$_1$ peak is a direct consequence of the two-component nature.

\begin{figure*}
\includegraphics[width=120mm]{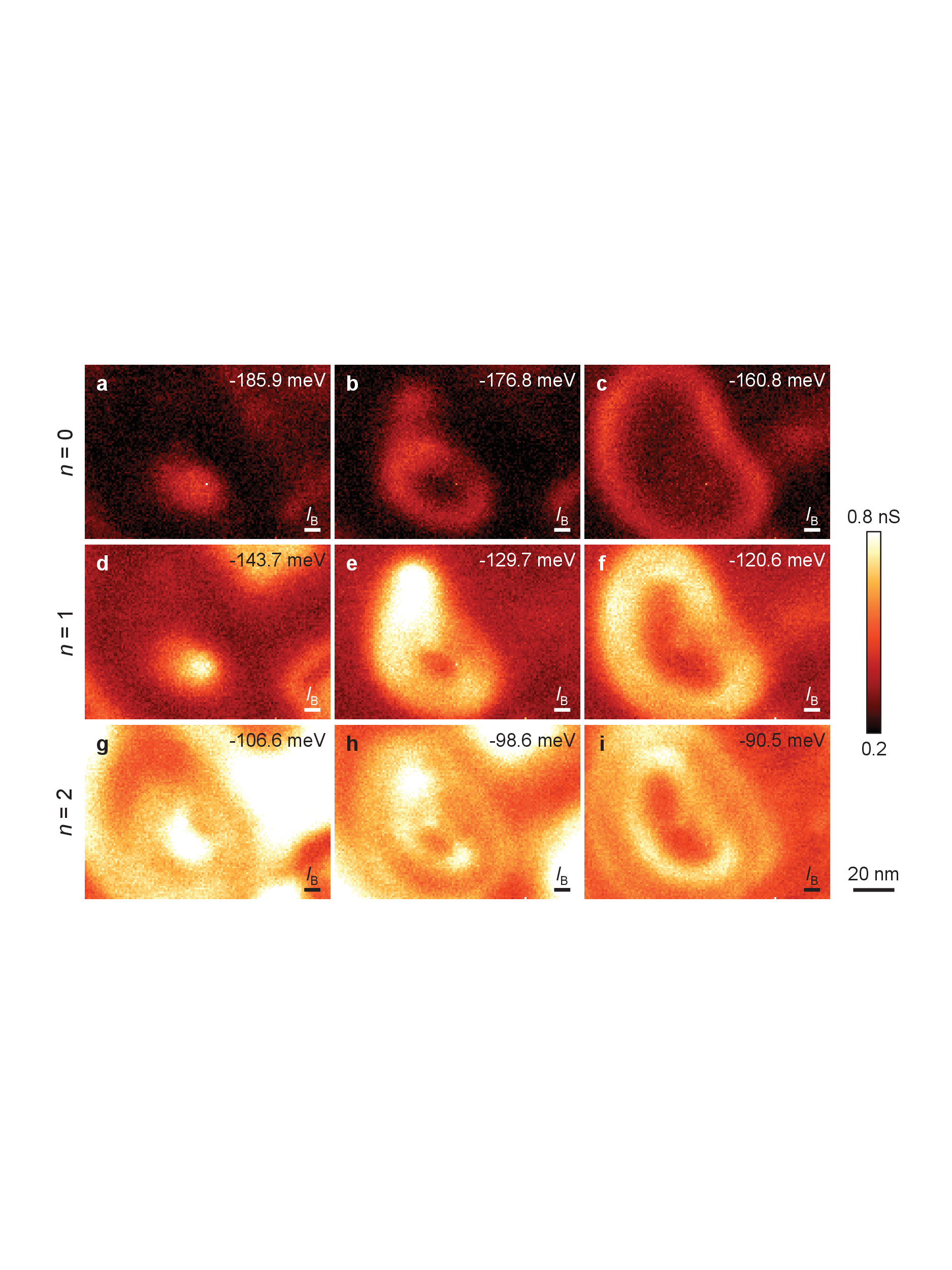}
\caption{
\textbf{Spatial and energy evolutions of localized Landau orbits trapped inside the potential dip at $\boldsymbol{B=}$11~T.}
Conductance images taken at different energies exhibit the ring-like trajectory of Landau orbits drifting along the equipotential lines.
\textbf{a-c},\textbf{d-f} and \textbf{g-i} are for LL$_0$, LL$_1$ and LL$_2$, respectively.
Complete data set is presented as a movie in the Supplementary Information.
The width of the ring gets wider and the concentric-ring structure becomes evident for LL$_2$.
The magnetic length $l_B$, which is a measure of the size of the LL$_0$ orbit, is shown in each panel for comparison.
}
\end{figure*}

Next we show the results of SI-STM around the potential minimum.
Figure~2 shows a series of conductance maps at 11~T in the same field of view of as in Fig.~1b-e.
All the maps exhibit prominent ring-like structures, which are ascribed to the Landau orbits drifting along the equipotential lines~\cite{Hashimoto2008PRL}.
The ring corresponding to the LL$_0$ state emerges at the potential minimum and expands with increasing energy (Fig.~2a-c).
With further increasing energy, the ring expands out of the field of view and 
another ring associated with the LL$_1$ state evolves (Fig.~2d-f).
Expansion of ring is also observed in the LL$_2$ state (Fig.~2g-i) and even higher LL$_n$ states (not shown).
The ring gets wider with increasing $n$ and splits into two concentric rings for LL$_2$, characterizing the internal structure of Landau orbits.

\begin{figure*}
\includegraphics[width=120mm]{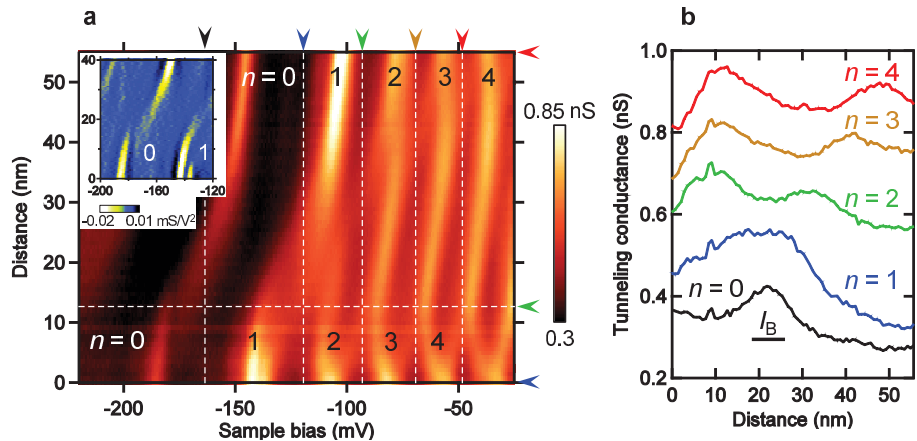}
\caption{
\textbf{Branching of Landau sub-bands and internal structures of Landau orbits.}
\textbf{a},
A false colour plot of the conductance-spectrum evolution from the potential minimum along the line shown in Fig.~1\textbf{b-e}.
Inset depicts the second derivative of conductance with respect to the bias voltage, which highlights the splitting features in the lower ($n=0$ and $n=1$ ) LLs.
Higher LLs evolve smoothly but split into two apparent branches.
Spectra shown in Fig~1\textbf{a} correspond to the horizontal line-cuts from this panel at distances marked by the horizontal arrows.
\textbf{b},
Vertical line-cuts from \textbf{a} at energies marked by the vertical arrows, showing internal structures of drifting Landau orbits for different $n$.
Each curve is offset vertically for clarity.
Although the number of peaks increases with $n$ in a conventional two-dimensional electron system (Ref.~\onlinecite{Hashimoto2012PRL}), there appear at most two peaks in the topological surface state of Bi$_2$Se$_3$.
The scale bar denotes $l_B$.
}
\end{figure*}

We further investigate the internal structure by analyzing a series of conductance spectra taken along the line shown in Fig.~1b-e.
As shown in Fig.~3a, the LDOS evolution shows the spatially dispersing Landau sub-bands.
Corresponding to the peak splitting shown in Fig.~1a, $n=0$ and $n=1$ Landau sub-bands are broken at the potential-gradient maximum and at the potential minimum, respectively.
The spatial evolutions of higher ($n>1$) Landau sub-bands are smooth but each sub-band broadens and splits into two apparent branches at the intermediate region, which correspond to the two concentric rings in Fig.~2.

We examined the detailed LDOS distribution across the drifting Landau orbits by taking vertical line-cuts from Fig.~3a (Fig.~3b).
As is already seen in Fig.~2, the Landau orbit gets wider with increasing $n$.
This behavior is common to both conventional~\cite{Hashimoto2012PRL} and Dirac~\cite{Okada2012PRL} systems, because the quantum Larmor radii for $n>0$ LLs, which characterize the widths of the Landau orbits, are given by $l_B\sqrt{2n+1}$ and $l_B\sqrt{2|n|}$ for conventional and Dirac systems, respectively (See Supplementary Information).

A remarkable difference between the two systems appears in the internal structures.
In the case of conventional systems, the LDOS variation across the drifting LL$_n$ orbit exhibits $n+1$ peaks because the corresponding wave function contains $n$ nodes~\cite{Hashimoto2012PRL,Yoshioka2007JPSJ} (See also Supplementary Information).
In contrast, in the case of the topological surface state of Bi$_2$Se$_3$, the number of peaks never exceeds two, even for $n>1$ LL states, as shown in Fig.~3b.

In the following, we show that our observations can be captured by model calculations and are direct consequences of the two-component wave function.
We adopt a model Hamiltonian $H=H_0+V(\mathbf{r})\sigma_0$, where $H_0$ represents the unperturbed Hamiltonian for two-dimensional Dirac electrons in $B$ and $V(\mathbf{r})$ is a circular-symmetric Coulomb potential generated by a subsurface charge.
$\sigma_0$ is the unit matrix.

\begin{figure*}
\includegraphics[width=160mm]{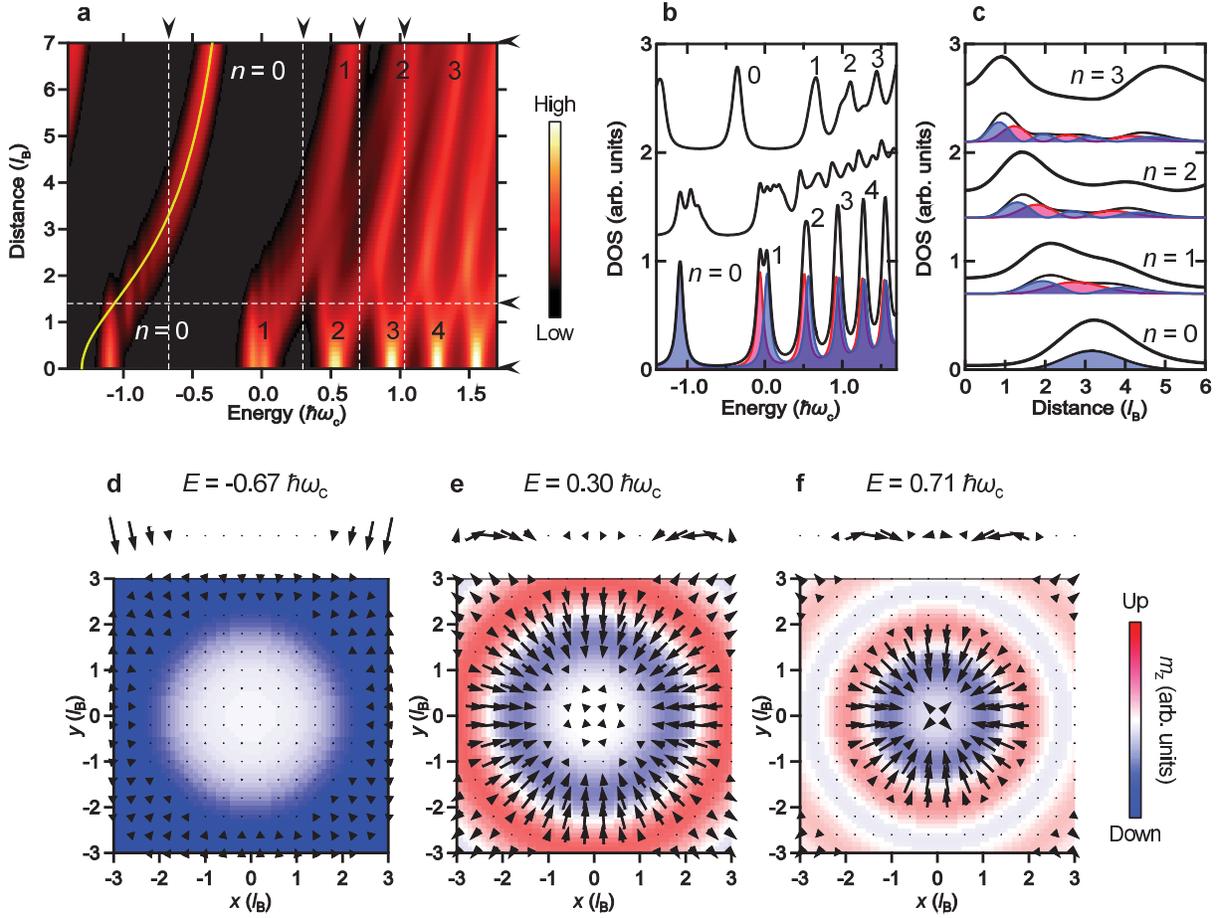}
\caption{
\textbf{Results of model calculations based on the two-component Dirac Hamiltonian}.
\textbf{a},
Intensity plot of calculated LDOS as a function of energy and distance from the bottom of the potential.
The yellow solid line denotes the radial variation of the potential used for the calculation.
The length is measured in units of $l_B$ and the energy is measured in units of $\hbar\omega_c$, where $\omega_c=\sqrt{2}v/l_B$ is the cyclotron frequency and $v$ is the electron velocity.
The damping parameter $\Gamma$ was set to $0.05\hbar\omega_c$.
See Supplementary Information for details.
\textbf{b},
LDOS spectra obtained by taking horizontal line-cuts at the representative points shown by horizontal arrows in \textbf{a}.
(From bottom to top, distance $|\mathbf{r}|=0$, $1.4\mathit{l_B}$ and $7.0\mathit{l_B}$, respectively.)
Each curve is offset vertically for clarity.
At the bottom of the potential, partial LDOS spectra associated with the up-spin (filled red curve) and down-spin (filled blue curve) components are also shown.
It is clear that LL$_0$ consists of down-spin component only and the splitting of the LL$_1$ peak is associated with the spin degrees of freedom.
\textbf{c},
Thick solid lines represent internal structures of Landau orbits obtained by taking vertical line-cuts at the representative energies shown by vertical arrows in \textbf{a}.
Partial LDOS (thin black lines) from the principle $j_z$ state and its up-spin (filled red curves) and down-spin (filled blue curves) components are also shown.
Data for each $n$ are offset vertically for clarity.
Nodes in the up-spin component are filled by the down-spin component and vice versa.
\textbf{d-f}
Spatial distribution of energy-dependent spin-magnetization vectors defined by $m_i=\frac{\hbar}{2}\sum_{n,j_z}\frac{\Gamma}{(E-E_{n,j_z})^2+\Gamma^2}\boldsymbol\Psi_{n,j_z}^\dagger(\mathbf{r}) \sigma_i \boldsymbol\Psi_{n,j_z}(\mathbf{r})$, where $\sigma_i(i=x,y,z)$ are Pauli matrices.
The in-plane components are indicated by arrows and the out-of-plane component is indicated by colours.
The line-cut at $y=0$ is also shown above each panel.
}
\end{figure*}

It should be noted that the good quantum number here is the total angular momentum $j_z$, which is a consequence of strong spin-orbit coupling.
This is in contrast to the case of conventional systems where the orbital angular momentum $l_z$ specifies the quantum states~\cite{Yoshioka2007JPSJ}.
Therefore, $H$ is block diagonalized with respect to $j_z$ and we can calculate the energy spectrum $E_{n,j_z}$, the wave function $\boldsymbol\Psi_{n,j_z}(\mathbf{r})$, and LDOS $D(E,\mathbf{r})=\sum_{n,j_z}\frac{\Gamma}{(E-E_{n,j_z})^2+\Gamma^2}|\boldsymbol\Psi_{n,j_z}(\mathbf{r})|^2$, assuming a Lorentzian broadening with a damping parameter $\Gamma$.
Details are given in the Supplementary Information.

Figure~4a shows an intensity plot of the calculated LDOS as a function of energy and $|\mathbf{r}|$, which reproduces the overall features of the experimental results shown in Fig.~3a.
The discrete vertical ridges seen in the $n=0$ Landau sub-band correspond to the different $j_z$ states, which are degenerate for $V(\mathbf{r})=0$.
Once $V(\mathbf{r})$ is turned on, this degeneracy is lifted because the Landau orbit with higher $j_z$ drifts at larger $|\mathbf{r}|$ where the potential energy is higher.

Figure~4b depicts the calculated LDOS spectra at representative points, resembling the observed tunneling spectra shown in Fig.~1a.
In particular, the splitting of the LL$_1$ peak at the bottom of the potential is well captured.
The physical picture of this splitting can be understood by looking into the nature of the wave function at $\mathbf{r}=0$.
By inspecting the functional form of $\boldsymbol\Psi_{n,j_z}(\mathbf{r})$ given in the Supplementary Information, one finds that $\boldsymbol\Psi_{n\neq0,j_z}(\mathbf{r}=0)$ consists of only two quantum states with $j_z=+1/2$ and -1/2, which originate from the up-spin and down-spin components, respectively.
Because these two states have different spatial extent, their energies are different; the LDOS peak splits accordingly.
Thus, the splitting of the LL$_1$ peak at $\mathbf{r}=0$ is a direct consequence of the two-component nature.
The splitting should also occur for LL$_n$ with $n>1$, but its detection is much harder because the energy difference between $j_z=\pm1/2$ states becomes smaller with increasing $n$.
Note that, the LL$_0$ peak does not split at $\mathbf{r}=0$ because only the down-spin component of $\boldsymbol\Psi_{0,j_z}(\mathbf{r}=0)$ is non-zero.
The relevance of this scenario is highlighted by looking at the spin-resolved LDOS at $\mathbf{r}=0$ (Fig.~4b).

The two-component nature also explains the absence of nodal structure in the LDOS distributions.
The $|\mathbf{r}|$ dependence of the calculated LDOS (Fig.~4c, thick black curves) exhibits only two peaks for $n>0$, being in agreement with the experiment.
We also plot the spin-resolved partial LDOS associated with the principle $j_z$ states which mainly contribute to the LDOS at a given energy.
Although the down-spin component (blue) has $n$ nodes as in the case of conventional systems, the number of nodes for the up-spin component (red) is $n-1$.
Therefore, the nodes for one component are always filled by the other and two enhanced LDOS peaks are formed near the edges.

The above discussion indicates that the potential variation not only affects the orbital motion but also induces non-trivial spin-magnetization textures through the strong spin-orbit coupling.
Indeed, calculated spin-magnetization distributions shown in Fig.~4d exhibit energy-dependent cycloidal-helix-like patterns along the radial direction.
The combination of Landau quantization and a tailored potential landscape may provide a novel "magnetoelectric" control of spin degrees of freedom.
For example, tip-induced local gating~\cite{Fu2013ACSNano} may be utilized to manipulate spin-magnetization textures.
We anticipate that this leads to intriguing spintronic and topological applications.

\noindent
\subsection*{Methods}

Bi$_{2}$Se$_{3}$ crystals were prepared by the melt-growth technique.
SI-STM experiments were performed at 1.5~K with a commercial low-temperature ultra-high-vacuum STM (Unisoku USM-1300) modified by ourselves~\cite{Hanaguri2006JPhys}.
The clean and flat surface was obtained by \textit{in-situ} cleaving at $\sim$77 K.
After the cleaving, the sample was transferred quickly to the STM unit which was kept below 10~K.
Magnetic field was applied perpendicular to the cleaved surface.
An electro-chemically etched tungsten wire was used as an STM tip, which was cleaned and characterized \textit{in-situ} with a field-ion microscope.
Tunneling spectra were taken with a software lock-in detector integrated in a commercial STM controller (Nanonis).

\subsection*{Acknowledgments}

The authors thank X. Chen, K. Iwaya, Y. Kohsaka, K. Nomura, M. Ogata, Y. Okada and A. W. Rost for discussions. This work was partly supported by Grant-in-Aid for Scientific Research from the Ministry of Education, Culture, Sports, Science and Technology of Japan (Grant No. 23103519).

\subsection*{Author contributions}

Y.S.F. carried out the experiments and M.K. performed the theoretical modeling and calculations.
Bi$_2$Se$_3$ single crystals were grown by K.I. and T.S.
T.H. and H.T. supervised the project. 
Y.S.F., M.K. and T.H. designed the experiment, analyzed the data and wrote the manuscript.

\subsection*{Competing financial interests}
The authors declare no competing financial interests.

\subsection*{Supplementary information}
Available at \url{http://www.riken.jp/epmrt/Hanaguri/SI/Bi2Se3_2comp/SI.html}

\end{document}